# Maximizing Smith-Purcell radiation intensity via multi-parameter optimization

Xiaojie Yang[1], Shichao Zhao[1], Xinyu Liu[1], Suguo Chen[2], Yue Wang[2], Ping Zhang[1], Zhigang Lu[1], Zhanliang Wang[1], Sunchao Huang[1*], and Yubin Gong[1†]

[1]National Key Laboratory of Science and Technology on Vacuum Electronics, School of Electronic Science and Engineering, University of Electronic Science and Technology of China
[2]Department of Applied Physics, Xian University of Technology, Xian 710048, China
[*]sunchaohuang@uestc.edu.cn; [†]ybgong@uestc.edu.cn

**Abstract:** Smith-Purcell radiation offers a versatile platform for terahertz light-matter interactions and compact source development, with intensity enhancement as a key priority. Using particle-in-cell simulations across a duty-cycle range of 0.1-0.9 at 0.4, 0.5, and 0.6 THz, we co-optimize the grating period, electron kinetic energy and duty-cycle. The intensity exhibits a unimodal dependence on the grating period at all three frequencies, with peak enhancements of 96-, 34-, and 17.5-fold relative to our reference baseline $I_0$. The optimal duty cycle shifts from 0.2 to 0.4 with increasing frequency, a trend arising from the trade-off between coherent superposition and beam-wave coupling under fixed-length conditions. These results provide systematic design guidelines for compact tabletop terahertz sources.

## 1. INTRODUCTION

When free electrons interact with periodic structures, dielectric interfaces, or external electromagnetic fields, they give rise to a variety of distinct electromagnetic radiation phenomena, among which Smith-Purcell radiation[1–6], transition radiation[7–10], and Cherenkov radiation[11–14] represent the three most extensively studied mechanisms. Free-electron radiation sources built upon these fundamental mechanisms have long been established as versatile platforms for generating high-power, widely tunable electromagnetic radiation, with a spectrum spanning from terahertz to X-rays[15–17]. Smith-Purcell radiation, first observed by Smith and Purcell in 1953, arises when an electron beam skims parallel above a periodic metallic grating[1,18,19]. Over the past decades, research on Smith-Purcell radiation has covered both fundamental mechanism studies and device-driven application developments. Owing to its salient advantages of compact footprint, high output power, and broad tunable bandwidth[20–23], Smith-Purcell radiation has demonstrated considerable potential for tabletop terahertz radiation sources and miniaturized vacuum electronic devices, drawing extensive attention across the vacuum electronics and terahertz science communities[14,24–26]. These properties also position Smith-Purcell radiation highly promising for cutting-edge applications including high-resolution imaging[27,28] and high-speed wireless communications[29,30]. However, low emission intensity remains a primary bottleneck limiting the development of Smith-Purcell radiation devices.

This bottleneck is largely attributed to a long-standing design ambiguity in Smith-Purcell light sources arises from the fact that multiple (period, energy) pairs can generate the same frequency, yet their influence on radiation intensity remains poorly understood. This ambiguity is evident from the dispersion relation[31–33]:

$$\lambda = \frac{d}{|n|}(\beta^{-1} - \cos\theta) \tag{1}$$

Where $\lambda$ is the radiation wavelength, $d$ is the grating period, $\beta = v/c$ is the reduced electron velocity, $\theta$ is the observation angle, and $n$ is the radiation order. This fundamental relationship has been widely verified across a broad spectral range through theoretical calculations and experimental measurements[1,34–36]. Based on this relation, numerous studies have realized broadband frequency tuning by adjusting electron beam energy, grating period, or observation angle, establishing mature frequency-control methodologies for Smith-Purcell radiation[37–41]. However, while frequency tuning is well established, achieving sufficient intensity remains the key challenge for practical applications. For a given target frequency, the achievable radiation intensity can vary by several orders of magnitude depending on the choice of structural parameters[42–44]. Prior studies have confirmed that the intensity of Smith-Purcell radiation is strongly dependent on both grating structural parameters and electron beam operating parameters, and that radiation efficiency can be substantially improved via targeted parameter tuning[45–48]. Nevertheless, most efforts to date have focused on individual parameters in isolation, and systematic exploration of multi-parameter co-optimization for maximum intensity remains limited. Therefore, identifying the optimal parameter combination that maximizes the radiation intensity is a critical next step.

Here, we resolve this design ambiguity by establishing a universal optimization paradigm for fixed-length grating structures that maximizes radiation intensity through optimal selection of grating period, electron kinetic energy, and duty cycle. Specifically, we perform comprehensive parameter-space scans across duty cycles ($w/d$) for three representative target frequencies: 0.4, 0.5, and 0.6 THz. By constructing complete intensity optimization profiles, we not only identify the globally optimal parameter combinations for each target frequency but also systematically reveal the evolution of radiation intensity and its underlying physical mechanism under multi-parameter interplay. The intensity exhibits a unimodal dependence on the grating period at all three frequencies, with peak enhancements of 96-, 34-, and 17.5-fold relative to our reference baseline.

## 2. RESULTS AND DISCUSSION

Fig. 1(a) presents the schematic diagram of the Smith-Purcell radiation configuration investigated in this work. The periodic metallic grating is characterized by four geometric parameters: grating period $d$, groove width $w$, groove depth $h = 150$ μm, and overall structural length $L = 6000$ μm. The duty cycle $w/d$ is systematically varied from 0.1 to 0.9 to quantitatively explore its modulation effect on radiation performance. The radiation intensity is detected at a distance $R = 14000$ μm from the grating center along the x-direction, with the observation angle set normal to the grating surface ($\theta = 90°$).

All numerical simulations are carried out via the particle-in-cell solver in CST, based on the finite integration technique for fully self-consistent beam-wave interaction calculations. The metallic grating is modeled as a perfect electric conductor, which serves as a high-fidelity approximation for high-conductivity metals in the terahertz regime.

The electron beam is set to have a Gaussian-profile beam, and the emitter surface height is set to 55 μm. The vertical distance between the electron-beam center and the grating surface is fixed at 32.5 μm, establishing a uniform baseline for beam-grating near-field coupling across all simulation cases. On this basis, to precisely lock the radiation frequency to the target value, the electron kinetic energy $E_{\mathrm{k}}$ is carefully determined for each grating period d to satisfy the fundamental-order ($|n| = 1$) Smith-Purcell radiation dispersion relation under normal observation: $\lambda = d/\beta$ . This constraint ensures that all sampling points yield radiation at the precise target frequency, such that the observed variation in radiation intensity arises solely from parametric optimization rather than frequency detuning. For spatial discretization, a globally adaptive hexahedral mesh is adopted to resolve the fine near-field structure around the grating. Open-boundary conditions are implemented in the x, y and z directions to suppress spurious boundary reflections that interfere with the radiation field.

The reference intensity $I_0$ is defined as the radiation intensity at a specific frequency for a grating period of $d = 100$ μm and a duty-cycle of $w/d = 0.7$. This unified benchmark ensures the quantitative comparability of intensity enhancement magnitudes across different frequencies, grating periods and duty cycles, and directly reflects the performance gain delivered by parametric optimization.

Taking the target frequency of 0.4 THz as an example, Fig. 1(b) presents the distribution of radiation intensity across different geometric parameters. This two-dimensional parametric map visually illustrates the continuous evolution of radiation performance with structural parameters, thereby enabling precise determination of the global optimum corresponding to maximum radiation intensity. Building upon this optimization process, Fig. 1(c) and 1(d) present the radiation spectra at two target frequencies of 0.4 and 0.5 THz, comparing the radiation performance of the parameter-optimized structures with that of the reference configuration. Specifically, the green dotted curve in each panel depicts the spectrum with the maximum radiation intensity achieved via parameter optimization at each target frequency, whereas the red curve denotes the reference spectrum selected as the intensity benchmark. The results demonstrate that the peak intensities of the optimized structures are enhanced by factors of 96 and 34, respectively, relative to their corresponding reference peak intensities.

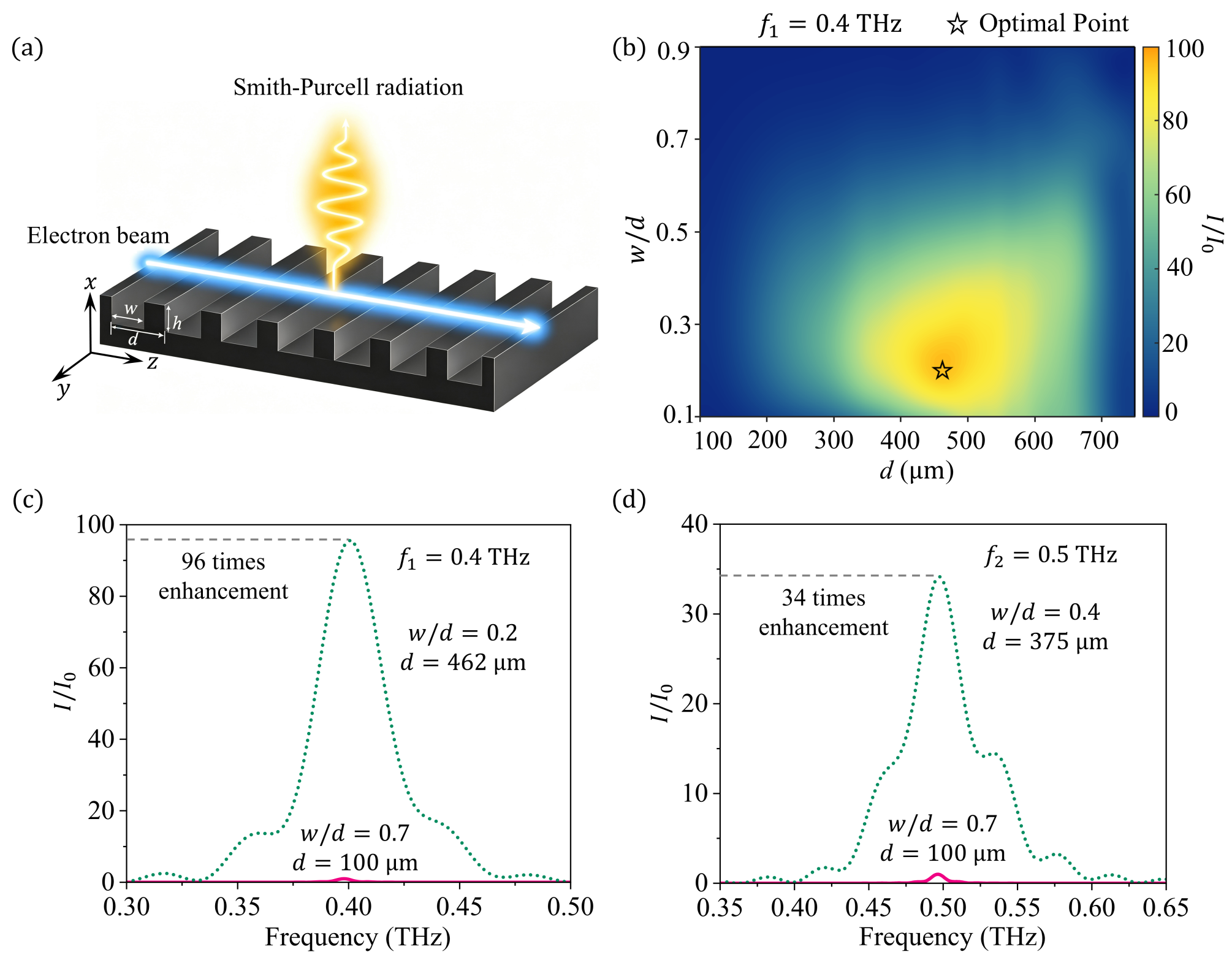

Fig. 1. **Significantly enhancing Smith-Purcell radiation intensity via multi-parameter co-optimization.** (a) Schematic diagram of the Smith-Purcell radiation configuration. (b). It illustrates the continuous evolution of radiation intensity during parameter optimization at the target frequency of 0.4 THz. The black star denotes the optimal point. Here, $I$ is the spectral intensity corresponding to different grating geometries, and $I_0$ is the reference intensity measured at $d = 100$ μm and $w/d = 0.7$. (c)–(d) Radiation spectra at the target frequencies of 0.4 THz and 0.5 THz, respectively. The peak radiation intensity is enhanced by a factor of 96 and 34, respectively, relative to the reference configuration ($d =$ 100 μm, $w/d = 0.7$).

Taking the 0.5 THz target frequency as a representative example, we first systematically present the parametric optimization results to determine the optimal combination of electron kinetic energy $E_k$ and grating period $d$ that maximizes the radiation intensity. For a fixed target frequency, each grating period $d$ corresponds to a unique electron kinetic energy $E_k$ satisfying the Smith-Purcell radiation phase-matching condition. Accordingly, Fig. 2 essentially depicts the evolution of radiation intensity as a function of the matched parameter pairs, rather than the independent effect of any single parameter.

The most striking feature is that parametric tuning enables an extremely wide tunable range of radiation intensity. Across all duty cycles, the radiation intensity enhancement factor exhibits a non-monotonic single-peak profile as a function of the matched parameter pairs. For small grating periods and their corresponding low electron energies, the intensity rises concurrently with increasing d and $E_{\mathrm{k}}$, and it decreases monotonically with further growth of both parameters after reaching a well-defined global maximum. The peak of each curve marks the globally optimal parameter pair. This result demonstrates that satisfying the phase-matching condition alone does not guarantee high radiation intensity. Only through the co-optimization of electron kinetic energy and grating period can the intensity limit be approached.

As shown in Fig. 2(a-e), both parameters exhibit a monotonically increasing trend in the low-to-medium duty cycle regime. At duty cycles of 0.1 and 0.2, the optimal parameter pair remains stable at $d = 316$ μm and $E_{\mathrm{k}} = 90.12$ keV. As the duty cycle increases to 0.5, the optimal grating period rises to 400 μm, with the corresponding electron kinetic energy increasing to 174.58 keV. The increase in duty cycle results in an increased groove width, which maintains efficient beam-wave coupling even at larger grating periods. Consequently, the optimum point constrained by phase matching shifts toward parameter combinations that yield higher radiation amplitude per period.

As shown in Fig. 2(f-i), in the high duty cycle range, parameter variation deviates from the monotonically increasing trend and exhibits non-monotonic fluctuation. The optimal grating period fluctuates in the range of 462 – 500 μm, and the corresponding electron energy varies from 289.88 to 413.43 keV. At a duty cycle of 0.9, the optimal parameter pair slightly decreases to $d = 470$ μm and $E_{\mathrm{k}} = 311.07$ keV. Under high duty cycles, grating ridges undergo a marked reduction in width, and the narrow ridges weaken the lateral confinement of the evanescent field, leading to lateral broadening of the near-field distribution. This not only breaks the monotonic growth of beam-wave coupling intensity with groove width, but also alters the radiation emission characteristics of a single grating unit. The original balance between phase matching constraints and coupling efficiency gains is disrupted, ultimately causing a slight nonlinear drift of the global optimal operating point.

For the 0.5 THz target frequency, the global maximum intensity enhancement factor reaches 34, occurring at a duty cycle of 0.4, with the corresponding optimal parameter combination of $d = 375$ μm and $E_{\mathrm{k}} = 143.6$ keV (Fig. 2(d)). Peak intensity varies non-monotonically with duty cycle, reaching its maximum at a medium duty cycle. As shown in Fig. 2, when the duty cycle is below 0.4, the beam-wave coupling enhancement effect dominates, and peak intensity rises with increasing duty cycle. Conversely, when the duty cycle exceeds 0.4, peak intensity gradually decreases due to the combined effects of fewer grating units caused by larger periods and an increased radiation attenuation rate.

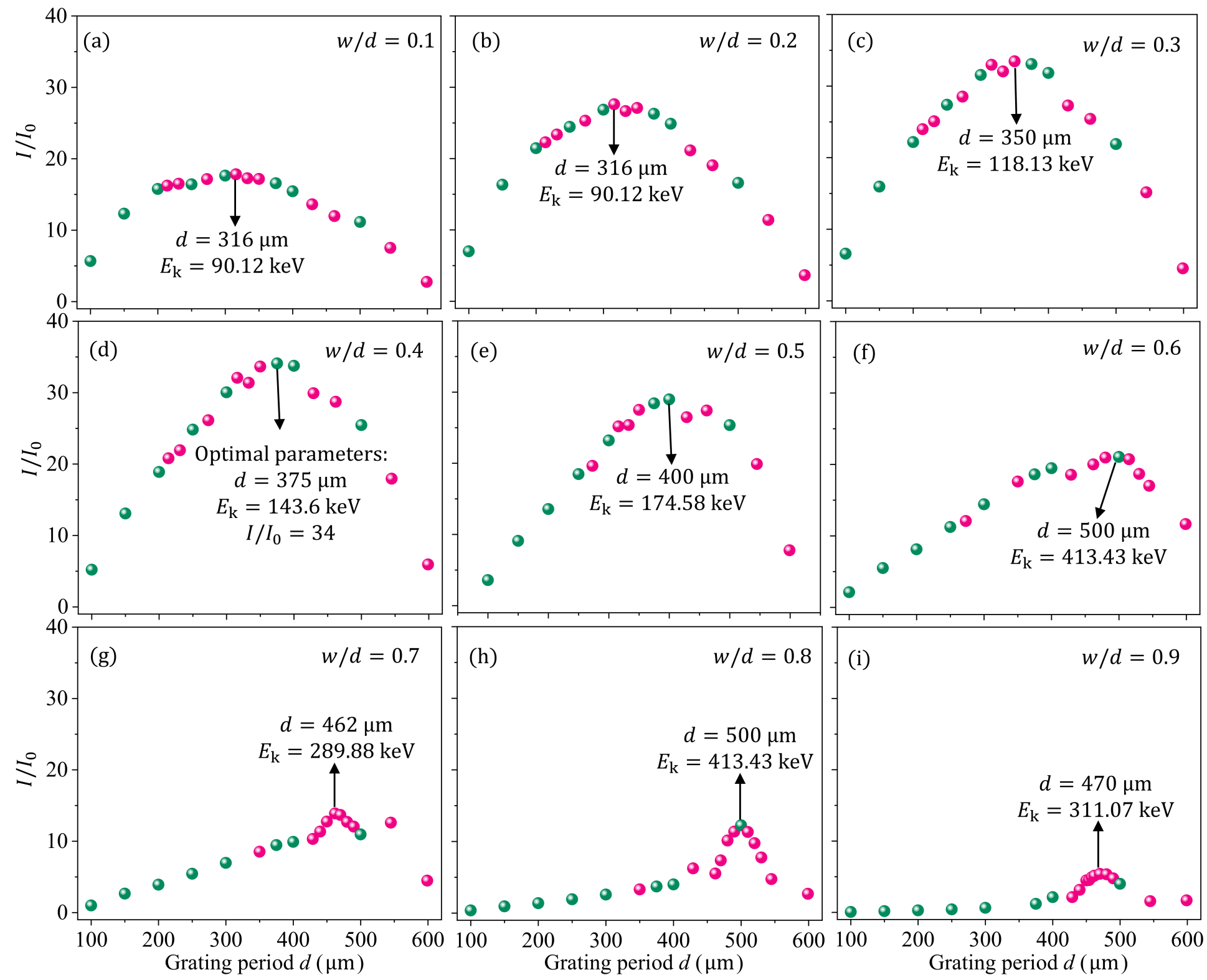

Fig. 2. **Parametric optimization of Smith-Purcell radiation for highest emission intensity at 0.5 THz.** (a)-(i) Normalized spectral intensity $I/I_0$ at the 0.5 THz target frequency as a function of grating period $d$ for duty cycles ranging from $w/d = 0.1$ to 0.9. Here, $I$ is the spectral intensity corresponding to different grating geometries, and $I_0$ is the reference intensity measured at $d = 100$ μm and $w/d = 0.7$. Green data points represent commensurate grating periods that can evenly divide the total structural length. Pink data points denote incommensurate periods. Notably, the global maximum intensity enhancement factor reaches 34 at $w/d = 0.4$ (panel d), yielding the optimal operating parameters of $d = 375$ μm and $E_k = 143.6$ keV for efficient 0.5THz Smith-Purcell radiation generation.

For the 0.4 THz target frequency, we employ the same optimization strategy as for 0.5 THz. For each grating period, the corresponding electron kinetic energy is precisely determined to ensure the radiation frequency matches the target, and the global optimum is finally identified via full parametric scanning.

The optimization behavior at 0.4 THz is highly consistent with that at 0.5 THz. As shown in Fig. 3, the radiation intensity enhancement factor exhibits a non-monotonic distribution with grating period at all duty cycles. The optimal parameter pair ($d$, $E_k$) also follows the two-stage evolution pattern of monotonic growth at low-to-medium duty cycles and non-monotonic fluctuation at high duty cycles as duty cycle increases. This result indicates that the trade-off mechanism between coherent superposition and beam-wave coupling under a fixed total length is consistent across both frequency bands. This consistency further verifies the effectiveness of the collaborative optimization method for grating period and electron kinetic energy in the low-frequency terahertz band.

The 0.4 THz band exhibits an extremely significant optimization gain. The global maximum enhancement factor reaches $I/I_0 = 96$ at a duty cycle of 0.2, corresponding to optimal parameters $d = 462\ \mu\text{m}$ and $E_\text{k} = 137.69\ \text{keV}$ (Fig. 3(b)). This means that without changing the total size, material, and structure of the grating, the 0.4 THz radiation intensity can be enhanced by nearly 100 times relative to the reference baseline, achieved solely by matching the optimal grating period and corresponding electron kinetic energy. Moreover, this maximum gain is approximately three times that obtained at 0.5 THz. Meanwhile, the optimal duty cycle for peak intensity shifts to a smaller value, which arises from the combined effects of beam-wave coupling efficiency and coherent superposition gain. At an extremely high duty cycle of $w/d = 0.9$, the radiation intensity curve flattens out with very low absolute intensity, and the optimization space is severely compressed.

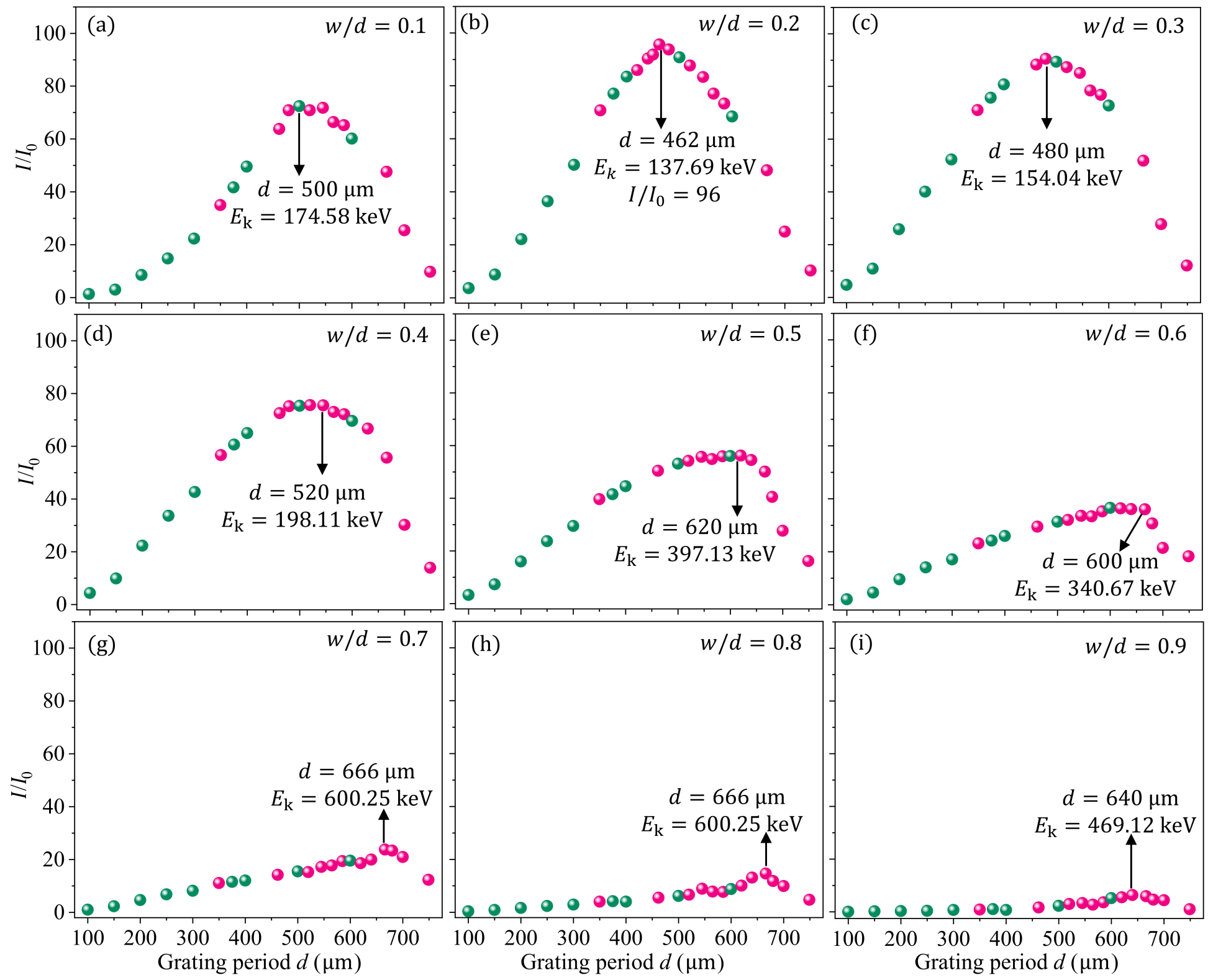

Fig. 3. **Parametric optimization of Smith-Purcell radiation for highest emission intensity at 0.4 THz.** (a)-(i) Normalized spectral intensity $I/I_0$ at the 0.4 THz target frequency as a function of grating period $d$ for duty cycles ranging from $w/d = 0.1$ to 0.9. Here, $I$ is the spectral intensity corresponding to different grating geometries, and $I_0$ is the reference intensity measured at $d = 100$ μm and $w/d = 0.7$. Green data points represent commensurate grating periods that can evenly divide the total structural length. Pink data points denote incommensurate periods. Notably, the global maximum intensity enhancement factor reaches 96 at $w/d = 0.2$ (panel b), yielding the optimal operating parameters of $d = 462$ μm and $E_k = 137.69$ keV for efficient 0.4 THz Smith-Purcell radiation generation.

We further extend the collaborative optimization method to 0.6 THz. By matching electron kinetic energy point-by-point to lock the target frequency, we obtain intensity optimization results across the full duty cycle range of 0.1 to 0.9.

The core characteristics including the unimodal variation of radiation intensity with grating period (rising first then falling) and the two-stage evolution of the optimal parameter pair ($d$, $E_\mathrm{k}$) with duty cycle are highly consistent with those at 0.4 and 0.5 THz. This further confirms that the trade-off mechanism between coherent superposition and beam-wave coupling under fixed total length is an inherent property of terahertz Smith-Purcell radiation, and the optimization method can cover the low-frequency THz band.

The optimization gain at 0.6 THz continues the trend of monotonic attenuation with increasing frequency. The global maximum enhancement factor $I/I_0 = 17.5$ appears at $w/d = 0.4$, corresponding to optimal parameters $d = 340$ μm and $E_\text{k} =$ 189.93 keV (Fig. 4(d)). Combining these with previous results, we observe that as the frequency increases from 0.4 THz to 0.6 THz, the maximum intensity enhancement factor gradually decreases from nearly 100-fold to just over 10-fold, while the optimal duty cycle shifts from 0.2 to stabilize at around 0.4. This evolution indicates that the equilibrium point between coherent superposition and beam-wave coupling gradually saturates at higher frequencies.

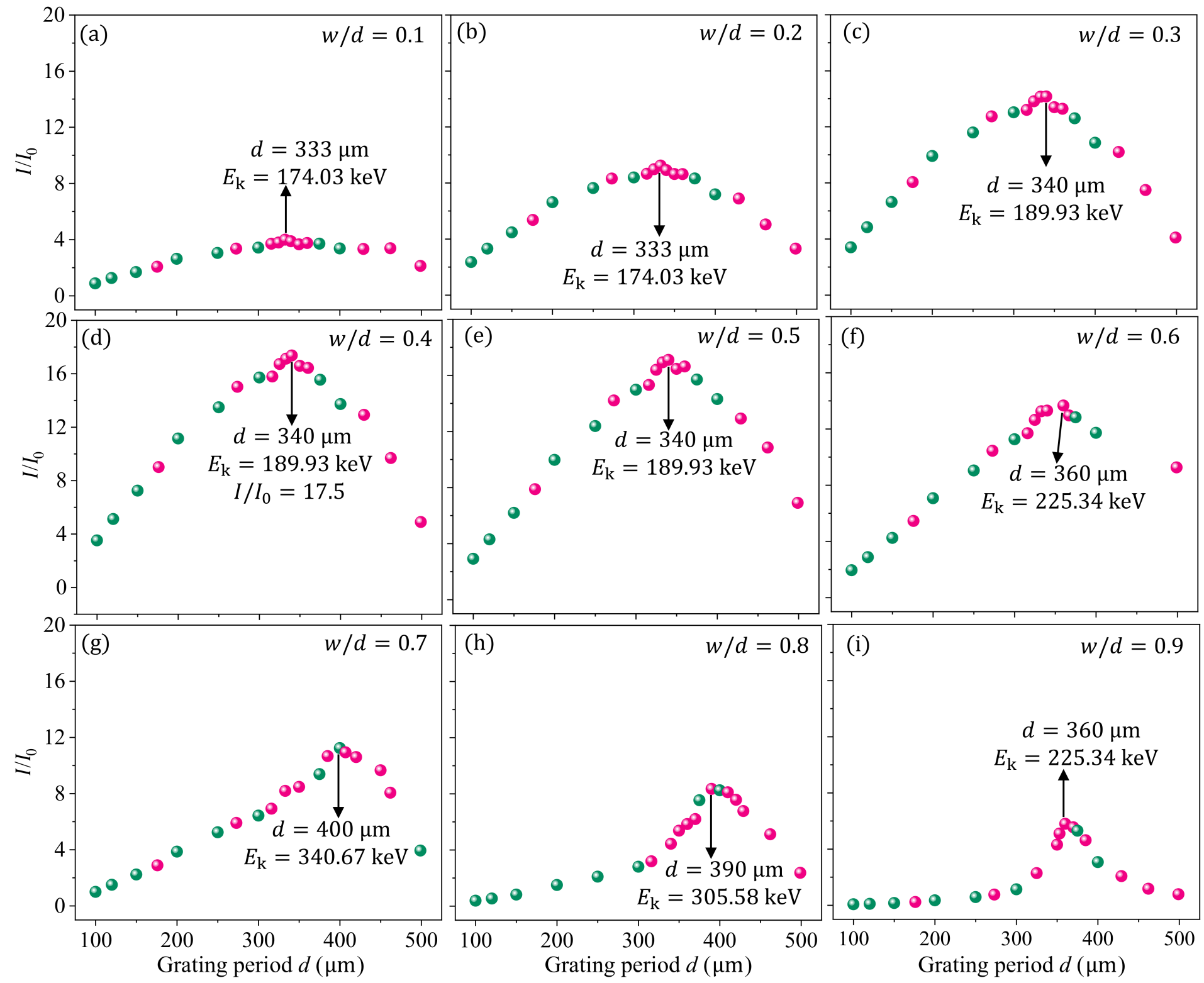

Fig. 4. **Parametric optimization of Smith-Purcell radiation for highest emission intensity at 0.6 THz.** (a)-(i) Normalized spectral intensity $I/I_0$ at the 0.6 THz target frequency as a function of grating period $d$ for duty cycles ranging from $w/d = 0.1$ to 0.9. Here, $I$ is the spectral intensity corresponding to different grating geometries, and $I_0$ is the reference intensity measured at $d = 100$ μm and $w/d =$ 0.7. Green data points represent commensurate grating periods that can evenly divide the total structural length. Pink data points denote incommensurate periods. Notably, the global maximum intensity enhancement factor reaches 17.5 at $w/d = 0.4$ (panel d), yielding the optimal operating parameters of $d = 340$ μm and $E_k = 189.93$ keV for efficient 0.6 THz Smith-Purcell radiation generation

In summary, we resolve a long-standing design ambiguity in Smith-Purcell light sources: for a target emission frequency, multiple (grating period, electron kinetic energy) combinations exist, yet their impact on radiation intensity remains underexplored. We establish a universal optimization paradigm for fixed-length grating structures that maximizes radiation intensity through optimal selection of these parameters. Without altering device dimensions or materials, we achieve enhancements of up to two orders of magnitude—a 96-fold increase at 0.4 THz—by systematically optimizing duty cycles across 0.1–0.9. Across 0.4–0.6 THz, we uncover a robust, frequency-invariant unimodal intensity–period relationship and a systematic optimal duty cycle shift from 0.2 to 0.4 with increasing frequency. This shift pinpoints a medium-duty-cycle sweet spot where coherent superposition and beam-wave coupling reach their optimal balance under fixed total length. Our multi-parameter strategy delivers a generalizable design rule that resolves the period-energy trade-off and provides a fabrication-free roadmap for upgrading compact terahertz sources.

**Acknowledgement**

This work was supported by National natural science foundation of China (Grants No.62527804, 62588201）

**Data Availability**

The data that support the findings of this study are available from the corresponding author upon reasonable request.